# Educational Customization by Homogenous Grouping of e-Learners based on their Learning Styles

Mohammadreza Amiri, Gholam Ali Montazer, Ebrahim Mousavi

{Mohammadreza.amiri, montazer, e.moosavi }@modares.ac.ir

Tarbiat Modares University

**Abstract**

The E-learning environment offers greater flexibility compared to face-to-face interactions, allowing for adapting educational content to meet learners' individual needs and abilities through personalization and customization of e-content and the educational process. Despite the advantages of this approach, customizing the learning environment can reduce the costs of tutoring systems for similar learners by utilizing the same content and process for co-like learning groups. Various indicators for grouping learners exist, but many of them are conceptual, uncertain, and subject to change over time.

In this article, we propose using the Felder-Silverman model, which is based on learning styles, to group similar learners. Additionally, we model the behaviors and actions of e-learners in a network environment using Fuzzy Set Theory (FST). After identifying the learning styles of the learners, co-like learning groups are formed, and each group receives adaptive content based on their preferences, needs, talents, and abilities. By comparing the results of the experimental and control groups, we determine the effectiveness of the proposed grouping method. In terms of "educational success," the weighted average score of the experimental group is 17.65 out of 20, while the control group achieves a score of 12.6 out of 20. Furthermore, the "educational satisfaction" of the experimental group is 67%, whereas the control group's satisfaction level is 37%.

**Keywords:** E-learning, Learning Style, Learner Behavior Modeling, Grouping, Customization of Learning, Fuzzy Set Theory.

# 1-Introduction

E-learning is an approach that aims to provide a well-planned, interactive, and learner-oriented learning environment for individuals, regardless of their location and time availability. It utilizes various digital technologies to create a flexible and distributed learning environment (Dovalli & Montazer, 2010). The key features of e-learning include extreme flexibility, learner-centeredness, and the elimination of time and place limitations, all of which contribute to creating equal opportunities for learning (Stephan et al., 1977).

In recognition of the diverse talents of learners, personalized learning has emerged as an approach to achieve effective learning by tailoring the learning content to the abilities and needs of each individual. This approach involves aligning teaching methods, educational programs, and content with the unique characteristics of each learner. Consequently, different versions of educational paths and teaching materials need to be created, taking into account the specific preferences of each learner. However, achieving this goal comes with challenges, such as the cost and time required to customize the curriculum and content for individual learners. To address these challenges, an alternative solution is to customize e-learning courses for groups of learners with similar characteristics within an e-learning network. By grouping learners based on their similarities, it becomes possible to create tailored educational experiences that are more efficient and cost-effective (Sanjabi & Montazer, 2020).

In summary, we are shifting the focus from the complex task of personalizing education and learning for each individual learner, which requires understanding their unique characteristics and designing tailored programs, to customizing the program and content for groups of learners who share similar characteristics Gomede, Miranda de Barros & de Souza Mendes, 2020). This approach allows us to streamline the process by tailoring the educational experience to a specific group rather than trying to cater to each individual learner separately.

To facilitate educational customization, we begin by identifying an index to group learners. Once this index is selected, learners are grouped together to receive personalized programs and lessons based on their individual talents. Subsequently, the outcomes of this approach are evaluated using indicators such as 'educational success' and 'educational satisfaction'.

One of the crucial elements in personalizing/customizing the e-learning environment is the concept of "learning style." A learning style refers to an individual's preferred way of comprehending, analyzing, and processing information during the learning process (Huang et al., 2020). It is influenced by learners' receptiveness to various information formats and their interaction with the learning environment. Understanding learners' learning styles is crucial for recognizing their unique differences in the educational process and promoting adaptability in e-learning environments (Dindar, Suorsa, Hermes, Karppinen & Näykki, 2021).

The objective of this article is to develop and implement an educational customization system. To achieve this, a method of homogeneous grouping is employed to categorize e-learners. Subsequently, adaptive content is presented to each group. The outcomes of the grouping process are evaluated using two criteria: educational success and satisfaction. Based on the mentioned points, the structure of the rest of the article is organized as follows: Section 2: Examining the concept of grouping and exploring different grouping methods. Section 3: Discussing various grouping indicators and introducing the concept of learning styles. Section 4: Presenting an explanation of the problem at hand. Section 5: Providing a brief introduction to fuzzy set theory and fuzzy expert systems. Section 6: Describing the design and implementation of the learning style fuzzy recognition system. Section 7: Explaining the process of grouping e-learners using the developed system. Section 8: Evaluating and analyzing the performance of the new customization system in a real application environment. Section 9: Concluding the article with a summary of the findings and implications.

## 2. The concept of grouping

The concept of grouping involves adapting the content and curriculum to meet the needs, abilities, and interests of learners, thereby enhancing the depth and quality of learning. To achieve this goal, it is essential to identify the factors that influence learning. By using this knowledge and implementing a well-designed educational approach, programs, and lessons can be customized to match the unique characteristics and abilities of learners, creating an engaging and interactive learning environment. Grouping e-learners entails dividing them into similar groups with the aim of attaining educational objectives, such as developing appropriate content and curriculum for each group (Montazer and Kamarei, 2018). The purpose of grouping is to enable learners to benefit from others by being part of specific groups, allowing them to interact with different learners, share ideas, and ultimately achieve improved learning outcomes (Jagadish, 2014). The grouping of learners can be accomplished through following four general methods (McGillicuddy and Devine, 2018; Nhan & Nhan, 2019; Jagadish, 2014).

I. Random grouping
II. Grouping chosen by the learners
III. Homogeneous grouping
IV. Heterogeneous grouping

### I. Random grouping

Random grouping involves the teacher randomly assigning learners to groups based on their own judgment. This method is commonly used in classrooms when there is a need to start teaching promptly and groups of equal size are required. It is considered a fair and unbiased approach, as each learner has an equal chance of being selected for any group. However, in random grouping, some learners may fall behind in the learning process as they may not be paired with suitable peers, deviating them from an effective learning path. Additionally, this method can hinder the dynamics of learning groups and make it challenging for learners to interact with one another (Nhan & Nhan, 2019).

### ii. Grouping chosen by the learners

In this method, learners themselves form the groups, and it is generally well-received by the learners. Typically, individuals who opt for this method are those who have previously succeeded in their learning experiences or have established social or academic connections with each other (Jagadish, 2014; McGillicuddy and Devine, 2018). Performance studies have shown that self-selected groups outperform random groups. The familiarity among members in self-selected groups facilitates communication and interaction. Learning in these groups creates a positive atmosphere among group members, leading to a favorable view of their peers. Members of self-selected groups have increased confidence in each other's abilities and appreciate the collaborative learning experience (Jagadish, 2014; McGillicuddyand Devine, 2018.

### iii. Homogeneous grouping

In this method, learners are grouped together based on similarities in their individual characteristics, such as personality, behavior, and talent (Huy, 2019). Homogeneous grouping aims to create groups of learners who share similar teaching and learning methods. In this approach, there is a higher degree of similarity among group members compared to other groups (Triantafillou et al., 2003). Studies have indicated that placing learners with higher abilities in homogeneous groups can enhance their learning speed (Franki lovi 2020). However, this type of grouping may give rise to potential issues, such as future group superiority within society (Huy, 2019).

**iv. Heterogeneous grouping**

In this approach, learners are grouped based on differences in personality, behavior, talent, and ability (Knisely, Levine, Vaughn-Cooke, Wagner & Fink, 2021). The purpose of this grouping is to create balanced groups comprising individuals with diverse abilities, skills, knowledge, gender, race, ethnicity, and other criteria. The underlying idea of this method is that each learner benefits from having a study partner within the group (Brame & Biel, 2015). In this method, capable learners gain a deeper understanding of their thinking processes, enhance their analysis of lesson content, and achieve better learning outcomes. Simultaneously, less capable learners benefit from the presence of more capable peers, thereby enhancing their own learning (Yulyng et al., 2021)

## 3. Grouping Indicators

Various indices have been employed to address the grouping problem in adaptive learning networks. Instead of relying on a single index, it is possible to integrate multiple grouping indices. The primary models for grouping e-learners based on similarity can be categorized into 13 different categories, including personality models, behavioral models, developed Personality models, Behavioral models, Developed personality models, Learning style models, Motivation, Feelings and emotions, IQ, Learning speed, Previous knowledge, Age and gender, Goal orientation, Memory and Self-efficiency. ().

In this article, the learning style index has been chosen from the aforementioned indicators for grouping learners. Therefore, we will provide further explanation about it in the following sections.

### 3-1. Learning Style

Learning style refers to the preferred approach an individual has when it comes to learning. It identifies the learner's preferences in terms of how they prefer learning content to be presented, how they interact with course materials, and how they absorb information from their environment (Hawk & Shah, 2007).

The Felder Silverman's Learning Style Model (FSLSM) has been widely utilized for personalizing e-learning environments and quantitatively assessing learning styles. This model categorizes learners into four dimensions: Perception, Input, Processing, and Understanding (Graf, Viola, Leo & Kinshuk, 2007; Nafea, Siewe & He, 2019).

In the "Perception" dimension, learners are classified into two categories: emotional and perceptive. Emotional learners prefer to acquire information through direct interaction with the real world, while perceptive learners prefer to gather information indirectly through their own conjectures and imagination.

In the "Input" dimension, visual learners have a preference for learning materials that are presented visually, such as figures, tables, and graphs. On the other hand, auditory learners prefer to receive information through auditory channels, such as listening to lectures or discussions.

In the "Processing" dimension, active learners thrive in interactive and cooperative learning environments, where they actively engage with other learners. Reflective learners, on the other hand, prefer to absorb information individually or in small groups, reflecting on it before further engagement.

The final dimension of learning style is "Understanding," which distinguishes between sequential and holistic learners. Sequential learners prefer to learn information in a linear, step-by-step manner, building upon each previous piece of knowledge. Holistic learners, on the other hand, seek a comprehensive understanding of the problem at hand, focusing on the bigger picture and generalities. These dimensions provide insights into how individuals prefer to perceive, receive, process, and understand information, contributing to their overall learning style.

Table 1 shows the different dimensions of the Felder-Silverman learning style.

**Table 1.** Dimensions and scales of Felder-Silverman's learning style (Carmona, Castillo, & Millán, 2008)

| Scale | Dimension |
|---|---|
| emotional/perceptual | Perception |
| visual/auditory | Input (Entrance0 |
| active/reflexive | Process |
| sequential/holistic | Understand |

In Table 2, educational preferences for different dimensions of Felder-Silverman's learning style are specified.

**Table 2.** The educational preferences based on learning style (García, Amandi, Schiaffino, & Campo, 2007)

| Dimension | Values | Educational preferences |
|---|---|---|
| Perception | Emotional | She prefers real examples; she should know all the working steps accurately and in detail. |
| | Perceptual | Prefers concepts and theories and thinks about the subject more than practical things. |
| Input | Visual | She prefers educational materials like pictures, diagrams, or videos. |
| | Verbal | She prefers teaching materials in text and audio. |
| Process | Reflexive | Learns things by experimenting, talking, and discussing with others helps them learn, prefer group study. |
| | Active | Demands are learned by thinking, and prefer individual study |
| Understanding | Sequential | Explaining the content part by part helps them, and providing a part-by-part structure and lesson topic is suitable. |
| | Overview | Expressing a general view of the subject helps them learn, and connection between different subjects helps to learn the subject matter. Expressing a summary of the lesson is effective in their learning. |

## 4. Problem Statement

In this article, the focus is on educational customization and grouping learners based on their learning style, specifically using the Felder-Silverman learning style index. The aim is to identify learners' learning styles indirectly through their network actions in an electronic learning environment. To account for the uncertainty and ambiguity in human behaviors and their interpretation, a mathematical model is developed using fuzzy set theory. This leads to the design and implementation of a fuzzy experience system.

By grouping learners based on their learning style, the educational strategy and adapted lesson content can be tailored to each group. The results obtained from this approach are then analyzed and evaluated. The overall goal is to provide a personalized learning experience by leveraging learners' individual learning styles and optimizing the educational process based on their needs and preferences.

## 5. Fuzzy set theory and fuzzy expert system

Fuzzy set theory, pioneered by Zadeh in 1965De Luca & Termini, 1993), recognizes that ambiguity and uncertainty are inherent in human thinking. It offers a mathematical framework to handle uncertain data and information, making it a crucial tool for modeling ambiguity. The fuzzy set theory employs linguistic variables and if-then rules to represent expert opinions and perform inference using fuzzy sets .

Fuzzy expert systems, which combine fuzzy set theory and fuzzy logic, provide a framework for modeling uncertainty in expert knowledge. They possess two main features:

1. Approximate Reasoning: Fuzzy expert systems are well-suited for approximate reasoning, especially in cases where deriving a mathematical model from a complex system is challenging. They allow for reasoning with imprecise and uncertain information.

2. Linguistic Variables: Fuzzy logic facilitates the use of linguistic variables, which are easily understood by humans. This enables the system to handle incomplete and uncertain information by representing it in linguistic terms.

In addition to these features, fuzzy expert systems consist of four key components:

**a) Fuzzifier**: The fuzzifier module defines the relationships between inputs and linguistic variables using membership functions. It converts the input variables into fuzzy numbers, representing the degree of membership in each linguistic variable.

**b) Knowledge base**: The knowledge base is created by combining the expertise of domain experts and formulating rules using linguistic variables. These rules express the relationships between input and output fuzzy sets. A typical fuzzy rule follows the format :

"If the input conditions are satisfied, then the output results can be deduced".

**c) Inference engine**: The inference engine is responsible for decision-making and inference using fuzzy rules and operators. One commonly used method is the Mamdani product inference engine, which applies fuzzy rules to determine the appropriate output based on the input variables.

**d) Defuzzifier**: The defuzzifier module reverses the fuzzification process by generating a definite output value from the fuzzy sets produced by the inference engine. It aims to obtain a crisp output that can be easily understood and utilized. Evaluating the stability of the fuzzy system is crucial, and certain conditions can lead to rule instability, such as conflicts with expert knowledge and opinions or similarities in the first part of rules but differences in the subsequent parts (Ghorbani & Montazer, 2015).

These components work together to enable the fuzzy expert system to process uncertain and imprecise information, make decisions based on fuzzy rules, and generate crisp outputs for further analysis or action.

## 6. Fuzzy learning style recognition system

In this part, according to the concepts of the previous part, the fuzzy experience system is designed to identify the learning styles of learners. Each of the components of this system is introduced below:

### 6-1. The input variables of the fuzzy system

The fuzzy learning style recognition system uses the behaviors of learners as input variables, which relate to different aspects of learning style according to experts. To collect this information, a questionnaire was created to identify the behaviors displayed by learners. Experts were asked to rate the relevance of each behavior to different dimensions of learning style using a range between very low, low, medium, high, and very high (Ghorbani & Montazer, 2015) and (Shiau, Wei, & Chen, 2015). The behaviors and their ratings can be found summarized in Table 3.

**Table 3.** The network behaviors corresponding to the learning style dimensions of the Felder-Silverman model Sanjabi & Montazer, 2020)

| Learning style dimension | Corresponding network behaviors |
|---|---|
| **Process** | • The amount of participation in class groups (the number of text messages sent and the number of audio and video contributions in the class)<br>• The amount of participation in the chat (number of text messages in the chat group)<br>• The amount of participation in troubleshooting sessions (number of text messages sent and audio and video contributions in troubleshooting sessions)<br>• Duration of the test<br>• Time devoted to training<br>• Number of related people in the class |
| **Perception** | • Time dedicated to theoretical and non-practical lessons<br>• Time allocated to the lesson<br>• Test time, the participation rate in troubleshooting sessions<br>• Number of studied examples<br>• Difficulty level of selected examples<br>• The number of requests from the system |
| **Entrance** | • Duration of listening to audio files<br>• Duration of using text lessons<br>• Duration of using video lessons, charts, and videos |
| **Understand** | • Number of studied examples<br>• Number of searches on related topics<br>• Time allocated to lesson summary |

Due to the utilization of vague and qualitative expressions such as "much," "very much," or "little" by experts when describing these behaviors, it becomes necessary to employ fuzzy sets for modeling these variables. Table 3 outlines the numerical ranges defining the linguistic variables associated with network behaviors. The membership function of these linguistic variables is expressed using fuzzy numbers, and trapezoidal numbers are preferred due to their flexibility (Ghorbani & Montazer, 2015). Each linguistic variable is denoted by an ordered quadruple (X, U, T, M), where X represents the variable (network behavior), U signifies the actual physical domain in which the linguistic variable X assumes its numerical values.

For instance, the behavior "participation in discussion groups" is represented by the number of contributions made by learners in the initial three sessions of the class. T represents the set of linguistic values, namely "high," "moderate," and "low" values, while M is the linguistic foundation that connects each linguistic rule in T to a fuzzy set in U. It is worth noting that the numerical intervals representing trapezoidal numbers for modeling learners' network behaviors were determined by sending the corresponding questionnaire to experts. The average opinion of the experts was then selected to define the numerical intervals.

The outcome of this modeling process is presented in Table 4.

**Table 4.** Fuzzy modeling of linguistic variables

| Behavior | Numeric range | Variable | Visual display |
|---|---|---|---|
| Participation in discussion groups | (0, 0, 3, 5) | Low | |
| | (3, 5, 8, 10) | medium | |
| | (8, 10, 15, 15) | Much | |
| Participation in chat | (0, 0, 3, 5) | Low | |
| | (3, 5, 8, 10) | medium | |
| | (8, 10, 15, 15) | Much | |
| Number of related people in the class | (0, 0, 1, 2) | Low | |
| | (1, 2, 3, 4) | medium | |
| | (3, 4, 5, 5) | Much | |

## 6-2. The output variables of the fuzzy system

The output variable of the fuzzy system corresponds to the score assigned to each learning style dimension of the Felder-Silverman model. This score is determined through a numerical questionnaire ranging from 0 to 11. A score closer to zero indicates a learning style that aligns more closely with the beginning of the spectrum. Table 5 provides the definition of these fuzzy variables.

**Table 5.** Fuzzy display of output variables

| Output | Numeric range | Visual display |
|---|---|---|
| Sensory | (0, 0, 6, 8) | |
| Sensory-intuitive | (6, 7, 8, 8) | |
| Intuitive | (6, 8, 12, 12) | |
| Consecutive | (0, 0, 6, 8) | |
| Sequential-General | (6, 7, 8, 8) | |

| | General | (6, 8, 12, 12) | 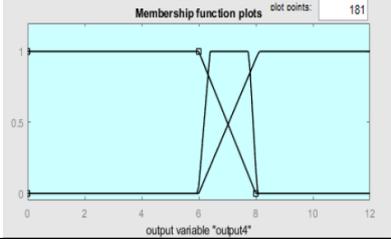 |
| | Visual | (0, 0, 6, 8) | 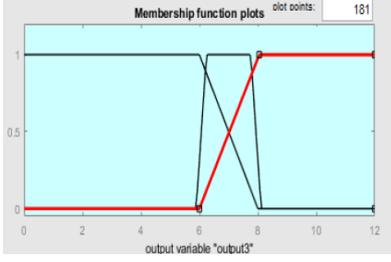 |
| | Visual-verbal | (6, 7, 8, 8) | |
| | Verbal | (6, 8, 12, 12) | |

### 6-3. Fuzzy learning style system rules

Fuzzy system rules are utilized to express the relationships between input and output variables. For instance, in the context of reflective learning style, learners exhibit a preference for group work, engage in communication with a majority of their peers in class, and tend to prioritize action over extensive contemplation. Based on the behaviors associated with the processing dimension, the aforementioned preferences can be captured by the following fuzzy rule:

Rule 1: If the level of participation in troubleshooting sessions is high, the level of participation in discussion groups is high, the number of individuals involved is high, the allocated time for testing is low, and the allocated time for practice is also low, then the learner is classified as reflective.

Rule 2: If the level of participation in troubleshooting sessions is low, the level of participation in discussion groups is low, the number of individuals involved is low, the allocated time for testing is high, and the allocated time for practice is high, then the learner is classified as reflective.

Rule 3: If the level of participation in troubleshooting sessions is average, the level of participation in discussion groups is low, the number of individuals involved is high, the allocated time for testing is low, and the allocated time for practice is average, then the learner is classified as reflective-reactive. Similar reasoning is applied to derive other rules within the system. Table 6 provides an overview of various network behaviors exhibited by learners, along with the linguistic variables and fuzzy rules associated with different dimensions of the learning style. Each column in this table represents a system rule.

**Table 6.** Some rules of learning style fuzzy identifying system

| Dimension | Linguistic variable values | | | | | Behavior |
|---|---|---|---|---|---|---|
| | Low | Low | Low | Low | Much | The amount of participation in discussion groups |
| | Low | Low | Low | Much | Much | The amount of participation in the chat |
| **Processing** | Low | Low | medium | Much | Much | The amount of participation in troubleshooting sessions |
| | Much | medium | medium | medium | Low | Time allocated to the test |
| | Much | medium | Low | medium | Low | Time devoted to training |
| | Low | medium | Much | Much | Much | The number of connected people |
| | **Reflection** | **Reflective-reactive** | **Reflective-reactive** | **Reactive-reflective** | **Reactive** | **Learning style** |
| **Perception** | Much | Much | Much | Low | Low | Time dedicated to theoretical and non-practical lessons |
| | Low | Low | Much | Low | Much | Real-time dedicated to lessons |
| | Low | medium | medium | Much | Much | Exam time |
| | Low | medium | Much | medium | Much | Number of studied examples |
| | Low | Low | Low | medium | Much | The difficulty level of selected examples |
| | Low | Low | Low | Much | Much | The number of requests from the system |
| | **Intuitive** | **Intuitive-Sensory** | **Intuitive-Sensory** | **Sensory-Intuitive** | **Sensory** | **Learning Style** |
| **Entrance** | Much | Much | Much | Low | Low | Duration of listening to audio files |
| | Much | Much | medium | medium | Low | Duration of using text lessons |
| | Low | medium | medium | medium | Much | Duration of using video |

| | | | | | lessons, charts, and videos |
|---|---|---|---|---|---|
| | **Verbal** | **Verbal-Visual** | **Visual-Verbal** | **Visual-Verbal** | **Visual** | **Learning Style** |
| **Understanding** | Much | Much | Low | Low | Low | Number of studied examples |
| | Much | medium | medium | Low | Low | Number of searches on related topics |
| | Much | Much | Much | medium | Low | Time allocated to lesson summary |
| | **Global-Sequential** | **Global-Consecutive** | **Global-Sequential** | **Sequential-Across** | **Consecutive** | **Learning Style** |

## 7. Implementation of learning style Identifier

In this study, the Elearning environment was created using the latest version of Moodle software, specifically Moodle 4.0.3 (Moodle Pty Ltd., 2022). This software facilitated the recording of various learner behaviors and captured their network actions during the initial three sessions of the course. The data collected from these interactions was then used to model learners' behaviors. The demographic information of the learner population is presented in Table 7.

Table 7. Demographic information of the statistical population

| Last educational certificate | | | Employment history (years) | | | Age | | | employment | | Gender | |
|---|---|---|---|---|---|---|---|---|---|---|---|---|
| MSC | BSC | Associate | More than 10 | 5-10 | Less than 5 | 30-35 | 25-30 | 20-25 | employed | student | F | M |
| 87 | 188 | 145 | 28 | 54 | 36 | 79 | 197 | 144 | 118 | 302 | 189 | 231 |

To ascertain the learning style of the learners, a comparison is made between the results obtained from the Felder-Silverman questionnaire and the modeling outcomes using the fuzzy system. Tables 8 and 9 present a juxtaposition of the questionnaire results and the fuzzy system outputs for the "input" and "processing" dimensions, respectively.

Table 8. Comparison of the results of the style identifier system and standard questionnaire in the entrance dimension

| Learner | type of learner | | Learning style scores | |
|---|---|---|---|---|
| | **Fuzzy system** | **Questionnaire** | **Fuzzy system** | **Questionnaire** |
| 1 | Visual | Visual | 9 | 8 |
| 2 | Visual | Visual | 8 | 10 |
| 3 | Visual | Visual | 9 | 10 |
| 4 | Visual- verbal | Visual | 6 | 9 |
| 5 | Verbal -visual | Visual | 5 | 8 |
| 6 | Visual- verbal | Visual | 6 | 10 |
| 7 | Verbal -visual | Visual | 5 | 9 |

Table 9. Comparison of the results of the style identifier system and standard questionnaire in the process dimension

| Learner | Type of learner | | Learning style number | |
|---|---|---|---|---|
| | Fuzzy system | Questionnaire | Fuzzy system | Questionnaire |
| 1 | Contemplative – reactive | Contemplative | 5 | 6 |
| 2 | Contemplative | Contemplative | 7 | 9 |
| 3 | Contemplative | Contemplative | 9 | 7 |
| 4 | Contemplative | Contemplative | 6 | 6 |
| 5 | reactive – Contemplative | Reactive | 4 | 2 |

The correlation coefficients between the results of the style identifier and the Felder-Silverman questionnaire are as follows: 86% in the "input" dimension, 78% in the "perception" dimension, 71% in the "processing" dimension, and 71% in the "understanding" dimension, respectively. Based on these values, it can be calculated that there is a strong correlation between the results obtained from the learning style dimensions identified through the questionnaire and the outcomes from the fuzzy recognition system. This correlation coefficient of 84% indicates a significant level of agreement between the two approaches.

**8. Grouping learners and customizing educational content according to each group**

By identifying the learning style of the learners, it becomes feasible to categorize them based on the similarity of their learning styles. The study's test population consisted of 420 individuals who had completed associate, bachelor, and master courses in various fields. Table 10 presents the characteristics of the Python programming language training course, while Table 11 provides the demographic information of the learner population.

Table 10. Details of the held training course

| Value | Characteristics |
|---|---|
| 13 | Number of sessions |
| Five weeks | Course duration |
| 420 | Number of participants |

The participants in this educational course had their learning styles determined using the style identification system, and subsequently, they were grouped into subgroups. Each subgroup received lesson content tailored to their learning style. The grouping process was based on the Felder-Silverman learning style index, which divides learners into four criteria: "Entry," "Perception," "Processing," and "Understanding." Based on the similarity of learners in each dimension, five distinct groups were formed as follows:

The first group: This group consists of 145 learners who exhibit the learning styles of "visual," "reflective," "sensory," and "holistic."

The second group: This group comprises 112 learners with the learning styles of "auditory," "reactive," "intuitive," and "sequential."

The third group: This group includes 104 learners who possess the learning styles of "visual," "reflective," "intuitive," and "sequential."

The fourth group: This group consists of 59 learners who have the learning styles of "auditory," "reactive," "sensory," and "holistic."

The fifth group: This group was formed with 46 participants who possess different characteristics and were selected as the "control" group, irrespective of their learning styles.

Following the grouping of learners based on their educational preferences related to each learning style (as shown in Table 2), customized lesson content was developed for each group. The effectiveness of this approach in enhancing the learning process and academic success of learners is depicted in Figure 1. The figure illustrates the positive impact achieved by appropriately grouping learners and delivering a curriculum tailored to the learning style of each group.

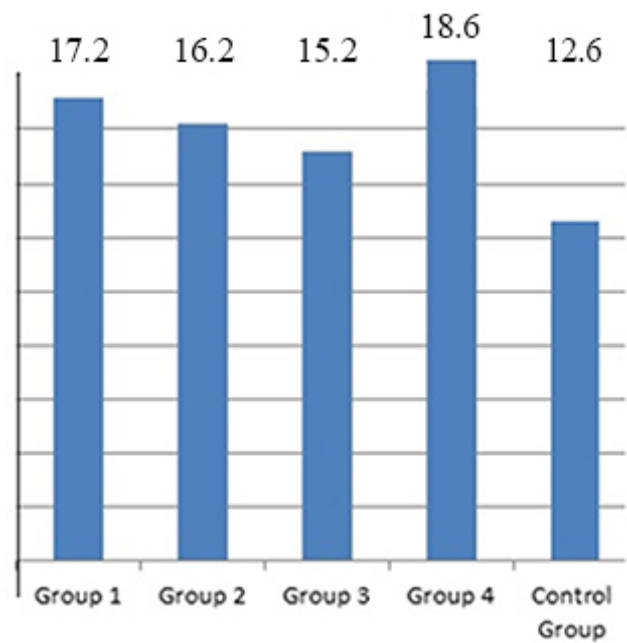

**Figure 1.** Impact of Tailored Learning Content on Academic Success Based on Learning Style Grouping

In this educational course, it is important to note that the "Python programming language" course was specifically chosen. The course spanned 13 sessions, which were conducted over a period of five weeks. To ensure a comprehensive and comparative content compilation, various formats for lesson presentations were incorporated, including images, tables, video files, audio files, and text files. The final exam for the course comprised of seven multiple-choice questions and three explanatory questions. It is worth mentioning that the allocated response time for the exam was consistent across all groups.

To evaluate the effectiveness of the grouping and customized training approach, two criteria were utilized: "academic success" and "academic satisfaction" (Rezaei & Montazer, 2016). Academic success was determined by analyzing the students' scores in the final test, with the average scores of different student groups compared to the average scores of the control group. Based on the results depicted in Figure 1, it is evident that grouping students into appropriate groups and providing tailored educational strategies based on their learning styles positively impacted the learning process and academic success of the students.

To check the level of academic satisfaction, the following questions were asked to the students:

1- How satisfied are you with the educational method of the system?

2. To what extent did you find the provided exercises (individual and group) favorable?

3. How accepting was you of the level of the exercises provided?

4- How useful was the feedback provided by the system?

5. To what extent did you find the provided content (audio/video/text) favorable?

6. How suitable was the communication method between you, other learners, and the teacher?

7. Overall, how satisfied are you with the system?

These questions aimed to capture the students' perceptions of their satisfaction with the educational method, exercises, feedback, content, communication, and overall satisfaction with the system.

Figure 2 shows the general architecture of the learning environment customization system:

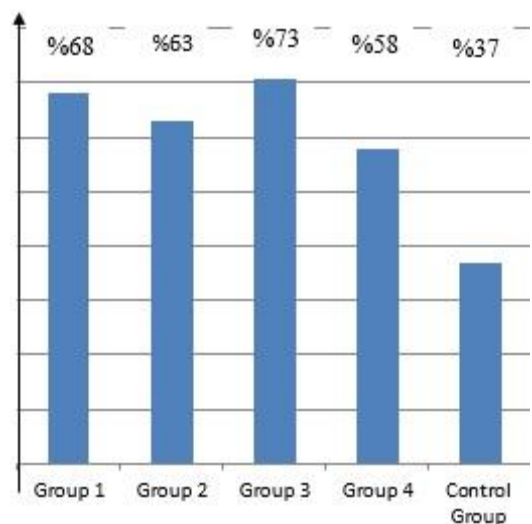

**Figure 2.** General Architecture of the Learning Environment Customization System

## 9. Evaluation of the customized environment

In order to assess and evaluate the outcomes of the customized learning environment, the "comparison of the average of two independent populations" test was employed. This test is widely utilized in statistical analysis to compare the means of two independent populations. Specifically, when the data follows a normal distribution, the t-test and associated statistics are suitable for comparing the means of two populations (Rochon, Gondan, & Kieser, 2012). By utilizing this test, statistical differences between the averages of two independent groups can be determined. It enables researchers to make informed decisions regarding the significance of the differences observed in the data.

### 9-1. Initial hypotheses for the implementation of t-test

Before conducting a two-sample t-test, it is important to consider the hypotheses for the data and test conditions. If these hypotheses are reasonably accurate, the two-sample t-test can be employed. However, if the data does not meet the assumptions of normality and equal variances, non-parametric methods should be utilized to compare the means of the two populations (Kim, 2015).

The t statistic is used to measure the average difference between two samples and is obtained by dividing the calculated difference by the standard error. The absolute value of t indicates the magnitude of the difference between the means. If the significance level of the difference between the average scores of the two groups is less than 5%, then the difference is considered significant. Conversely, if the significance level is greater than or equal to 5%, the difference between the average scores of the two groups is not statistically significant, leading to the rejection of the assumption that the two groups differ. The results of this test can be found in Table 11.

**Table 11.** t-test results for groups of recipients

| Group | The degree of significance of the difference | The amount of the t statistic | Significant Difference |
|---|---|---|---|
| Group 1 customized learning and control group | 2% | 2.38 | Positive |
| Group 2 customized learning and control group | 0% | 4.37 | Positive |
| Group 3 customized learning and control group | 2% | 2.69 | Positive |
| Group 4 customized learning and control group | 0% | 4.72 | Positive |

In Table 10, the column "significance of the difference between the average scores of two groups" indicates the significance level of the difference between the average scores of the two groups. If the value in this column is 0%, it signifies a statistically significant difference between the average scores of the two groups. This suggests that the outcomes of customized learning differ significantly from the normal and non-customized learning approach. On the other hand, if the value in this column is greater than 5%, it indicates that there is no statistically significant difference between the two evaluated groups.

Essentially, a significance level of 0% implies strong evidence to support the presence of a significant difference, while a significance level above 5% suggests that the observed difference is not statistically significant.

The one-way "analysis of variance" (ANOVA) method is utilized to compare the average scores among multiple groups. This method serves to assess the significance of both the one-way ANOVA itself and the disparities in average scores. It subsequently identifies the groups that exhibit a notable difference in average scores (Rezaei & Montazer, 2016) .The analysis reveals a significant result for the one-way analysis of variance, indicating that there are notable differences in the average scores among learners within the groups. Moreover, it is observed that there exists a significant difference in average scores between certain groups, with a significance level of 0.05. To assess the average differences between the groups, Table 14 is utilized.

Based on the findings from Table 15, it is observed that although there are differences in the average scores among groups one to seven, these differences are not statistically significant and can be disregarded. Consequently, it can be inferred that these seven groups, which have received tailored educational strategies based on their preferred learning styles, have exhibited commendable and acceptable learning efficiency. Their performance is largely comparable, suggesting that if the strategies and educational content of one group were applied to another group, their efficiency would likely decrease.

In contrast to groups one to seven, the control group displays a statistically significant and acceptable difference in average scores at a significance level of 0.05. This discrepancy can be attributed to the absence of a tailored learning strategy that meets the needs of the learners in the control group. The significant difference highlights the positive impact of program customization and curriculum content on academic success for the learners.

The analysis of the students' academic achievement test results at the conclusion of the training course clearly demonstrates the significant impact of grouping learners into appropriate groups and implementing tailored educational strategies based on their learning styles. This approach has proven to be highly influential in enhancing the learning process and promoting academic success among the learners.

## 10. Conclusion

In this article, we utilized the homogeneous grouping method, employing the Felder-Silverman learning style model, to achieve educational customization. To determine the learners' learning styles, we designed a fuzzy experience system that analyzes their actions within the electronic learning environment. This system takes the learner's behavior in the learning environment as input and produces their corresponding learning style as output. The system's rules establish the relationship between the learner's behavior and their learning style. Once the learning style was identified, learners were grouped together based on similarities in their learning styles. Each group then received personalized educational strategies tailored to their talents, preferences, and needs. To evaluate the effectiveness of our proposed method, we conducted an e-learning course. At the conclusion of the training course, we assessed its impact using two criteria: "academic success" and "academic satisfaction." The results demonstrated the efficacy of grouping learners based on their learning styles. The experimental group achieved a weighted average score of 17.65 (out of 20), while the control group scored 12.60. Furthermore, in terms of academic satisfaction, the average scores were 67% for the experimental group and 37% for the control group.